\documentclass[aps,prl,final,twocolumn,letterpaper]{revtex4}
\usepackage{graphicx}
\usepackage{epstopdf}
\usepackage{amsmath}
\usepackage{physics}
\usepackage{bm}
\usepackage{amssymb}
\usepackage{xcolor}
\usepackage{array}
\usepackage{multirow}
\usepackage{lipsum}
\usepackage[T1]{fontenc}
\usepackage{multirow}
\usepackage[english]{babel}
\usepackage{hyperref}

\renewcommand{\v}[1]{{\mathbf{\boldsymbol{#1}}}}

\usepackage{comment}
\usepackage{float}
\usepackage{tikz}
\usepackage{booktabs}
\usepackage{makecell}

\begin{document}

\title{Exact Valence-Bond Solid Scars in the Square-Lattice Heisenberg Model}
\author{David~D.~Dai}
\email{dddai@mit.edu}
\affiliation{Department of Physics, Massachusetts Institute of Technology, Cambridge, Massachusetts 02139, USA}

\date{\today}

\begin{abstract}
We show that the spin-$s$ square-lattice Heisenberg model has exact many-body scars.
These scars are simple valence-bond solids with exactly zero energy, and they exist in even-by-even systems and ladders of width $2$.
Ladders have additional scars corresponding to injecting one or two magnons on top of a parent valence-bond solid scar.
These scars have a remarkably simple physical origin based only the angular momentum algebra and cancellations from spin-antialignment within a valence bond.
Our comprehensive exact diagonalization calculations suggest that our valence-bond solids exhaust all exact eigenstates in the Heisenberg model except for few-magnon states near the top of the spectrum.
Our scars are interesting because they are not part of a tower, have area-law entanglement, break translation symmetry, and exist for Heisenberg models of all spin.
\end{abstract}

\pagestyle{myheadings}
\thispagestyle{empty}
\maketitle

\section{Introduction}

Recent experiments \cite{scar_first_exp} in quantum simulators and subsequent theoretical work \cite{PXP_theory_1, PXP_theory_2} have shown that an exciting new world lies high above the ground state in quantum many-body systems.
Contrary to the strong eigenstate thermalization hypothesis \cite{ETH1, ETH2, ETH3}, some nonintegrable systems including the PXP \cite{PXP_theory_1, PXP_theory_2, PXP_exact, PXP_su2, PXP_orbit}, AKLT \cite{AKLT_exact, AKLT_entanglement, AKLT_unified, MPS_scar}, spin-$1$ XY \cite{XY_1st, AKLT_unified, XY_EP}, and extended Hubbard models \cite{eta_pairing_1, Hubbard_SO4, eta_entanglement, extended_eta_scar} host a small number of many-body scar eigenstates, which have finite energy density yet are far from the corresponding Gibbs state.
These many-body scars lead to nonthermal dynamics starting from certain simple initial states \cite{scar_first_exp, PXP_theory_1, PXP_theory_2} and often have ground state-like properties including subvolume entanglement entropy and long-range order \cite{PXP_exact, AKLT_entanglement, XY_1st, eta_entanglement}.

We show analytically that the spin-$s$ square-lattice Heisenberg model, one of the most celebrated quantum antiferromagnets, hosts many-body scars for any $s$.
These scars are simple valence-bond solids with exactly zero energy, and they exist in even-by-even systems and ladders of width $2$.
Ladders have additional scars corresponding to injecting one or two magnons on top of a parent valence-bond solid scar.
Our scars have a remarkably simple physical origin based only the angular momentum algebra and cancellations from spin-antialignment within each valence bond.
Our comprehensive exact diagonalization calculations show that our valence-bond solids exhaust all exact states in the Heisenberg model except for few-magnon states near the top of the spectrum.
Compared to other many-body scars, ours are interesting in that they are appear to evade description by both the spectrum generating algebra and projector embedding frameworks. Our scars are also have area-law entanglement, break translation symmetry, and exist for Heisenberg models of all spin. 

\begin{figure}[t!]
  \centering
  \includegraphics[width=0.45\textwidth]{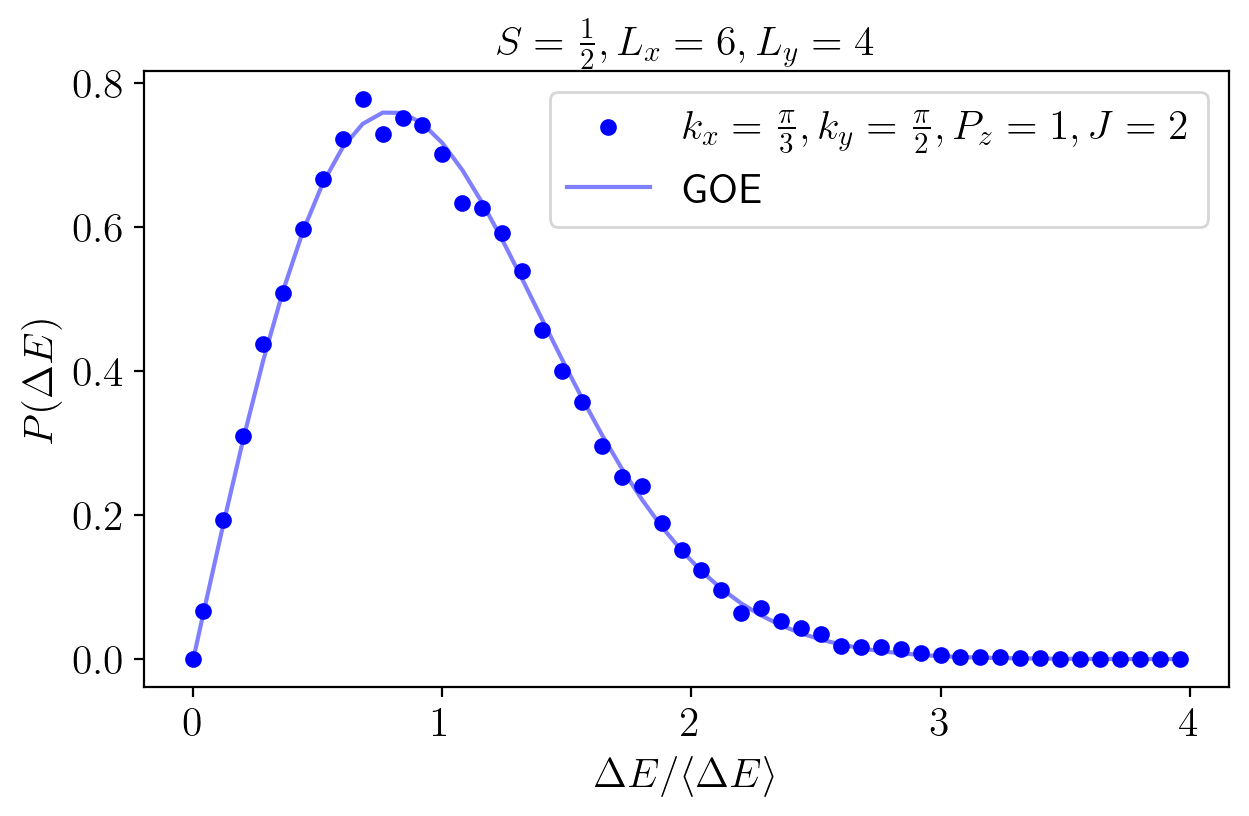}
  \caption{Level statistics for an arbitrary symmetry sector of the square-lattice Heisenberg model, showing good agreement with the Gaussian Orthogonal Ensemble (GOE).}
  \label{fig:LevelStatistics}
\end{figure}

The Hamiltonian of the square-lattice Heisenberg model is
\begin{equation}
    H =  \sum_{x=0}^{L_x - 1}\sum_{y=0}^{L_y - 1}
    \left( 
    \v S_{x,y} \cdot \v S_{x+1,y} + \v S_{x,y} \cdot \v S_{x,y + 1}
    \right),
\end{equation}
where $\v S_{x, y}$ is the spin operator on site $(x, y)$. We impose periodic boundary conditions $(x + L_x, y) \equiv (x, y)$ and $(x, y + L_y) \equiv (x, y)$. The square-lattice Heisenberg model is nonintegrable, which is confirmed by level statistics shown in Fig. \ref{fig:LevelStatistics}.  To the best of our knowledge, no exact eigenstates\footnote{We mean an eigenstate with an energy that is a simple rational number up to machine precision.} besides the trivial ferromagnetic and ferromagnetic plus few-magnon states were known in the square-lattice Heisenberg model before our work.

The model has several geometric symmetries: $x$-translation $T_x(x,y) = (x+1,y)$, $y$-translation $T_y(x,y) = (x,y+1)$, $x$-inversion $I_x(x, y) = (-x, y)$, and $y$-inversion $I_y(x, y) = (x, -y)$. $I_\mu$ and $T_\mu$ (we use Greek letters throughout to index directions) are commute only when $k_\mu \in \{0, \pi\}$ because $I_\mu T_\mu I_\mu = T_\mu^\dagger$.  A finite-size system also has rotation symmetry if $L_x = L_y$. From the spin algebra, we also have the spin-flip $P_z S_{x, y}^z P_z = - S_{x,y}^z$ and $\text{SU}(2)$ $J_z$ and $J^2$ symmetries. $P_z$ and $J_z$ commute only when $J_z = 0$.

\section{Exact Valence-Bond Solid Scars}

We start with a Heisenberg model on a single square to build intuition. Labeling the square's vertices counterclockwise as $1, 2, 3$ and $4$, the Hamiltonian is
\begin{equation}
    H = \v S_1 \cdot \v S_2 +
    \v S_2 \cdot \v S_3 + 
    \v S_3 \cdot \v S_4 + 
    \v S_4 \cdot \v S_1.
\end{equation}
The key insight is that the sum over all of a square's bonds can be factorized as
\begin{equation}
    \begin{aligned}
         H &= (\v S_1 + \v S_3) \cdot (\v S_2 + \v S_4)\\
         &= \frac{(\v S_1 + \v S_2 + \v S_3 + \v S_4)^2 - (\v S_1 + \v S_3)^2 - (\v S_2 + \v S_4)^2}{2},
    \end{aligned}
\end{equation}
enabling $H$'s eigenstates to be completely labeled by the quantum numbers $J_{1234},J_{13},J_{24},J_{1234}^z$.

We can interpret $J_{1234}=0,J_{13}=0,J_{24}=0$ as a valence-bond solid with spin-singlets on sites $13$ and $24$, $J_{1234}=1,J_{13}=1,J_{24}=0$ as a valence-bond solid with a spin-$1$ magnon on sites $13$, and $J_{1234}=0,J_{13}=1,J_{24}=1$ as a valence-bond solid with two spin-$1$ magnons together forming a spin-singlet bound state. Remarkably, all three of these states generalize to larger systems as exact mid-spectrum eigenstates.

\subsection{Ladders}
We now consider Heisenberg ladders of width $2$ and length $L$.
Let $s$ be the total spin of each site; our results apply to all $s$.
We impose periodic boundary conditions $(x + L, y) \equiv (x, y)$ and $(x, y + 2) \equiv (x, y)$.
The Hamiltonian can be factorized as
\begin{equation}
    \label{eq:ladder_H}
    \begin{aligned}
        H &= \sum_{x=0}^{L-1}\sum_{y=0}^1 \left( \v{S}_{x,y} \cdot \v{S}_{x+1,y} + \v{S}_{x,y} \cdot \v{S}_{x,y+1} \right)\\
        &= \sum_{x=0}^{L-1} \v{S}_{(x+1,0)(x,1)} \cdot \v S_{(x,0)(x+1,1)},
    \end{aligned}
\end{equation}
where $\v S_{(x+1,0)(x,0)} = \v S_{x+1,0} + \v S_{x,0}$. For consistency with larger systems, we count bonds $(x,y)-(x,y+1)$ and $(x,y+1)-(x,y+2)$ as distinct, even though the two connect the same sites.
This is equivalent to one double-strength bond $(x,y)-(x,y+1)$.

An exact zero-energy eigenstate is the diagonal valence-bond solid
\begin{equation}
    \ket{S_0} = \bigotimes_{x=0}^{L-1} \ket{\Phi}_{\left(x, 0\right)\left(x+1,1\right)}.
\end{equation}
Above, $\ket{\Phi}_{ij}$ is the unique spin-$0$ singlet on sites $i$ and $j$.
It satisfies
\begin{equation}
    \begin{aligned}
        \ket{\Phi}_{ij} &=
        \frac{1}{\sqrt{2s+1}}
        \sum_{k=0}^{2s} (-1)^{k}
        \ket{s,s-k}_i
        \otimes
        \ket{s,-s+k}_j,\\
        \v S_i \ket{\Phi}_{ij}
        &= -\v S_j \ket{\Phi}_{ij},\\
        P_z \ket{\Phi}_{ij}
        &= (-1)^{2s} \ket{\Phi}_{ij},\\
        \ket{\Phi}_{ji}
        &= (-1)^{2s} \ket{\Phi}_{ij}.\\
    \end{aligned}
\end{equation}
From the valence bonds, we have
\begin{equation}
    \v S_{(x,0)(x+1,1)} \ket{S_0} = 0,
\end{equation}
which combined with Eq. \ref{eq:ladder_H} proves that $H\ket{S_0} = 0$.

\begin{figure}[t!]
  \centering
  \includegraphics[width=0.225\textwidth]{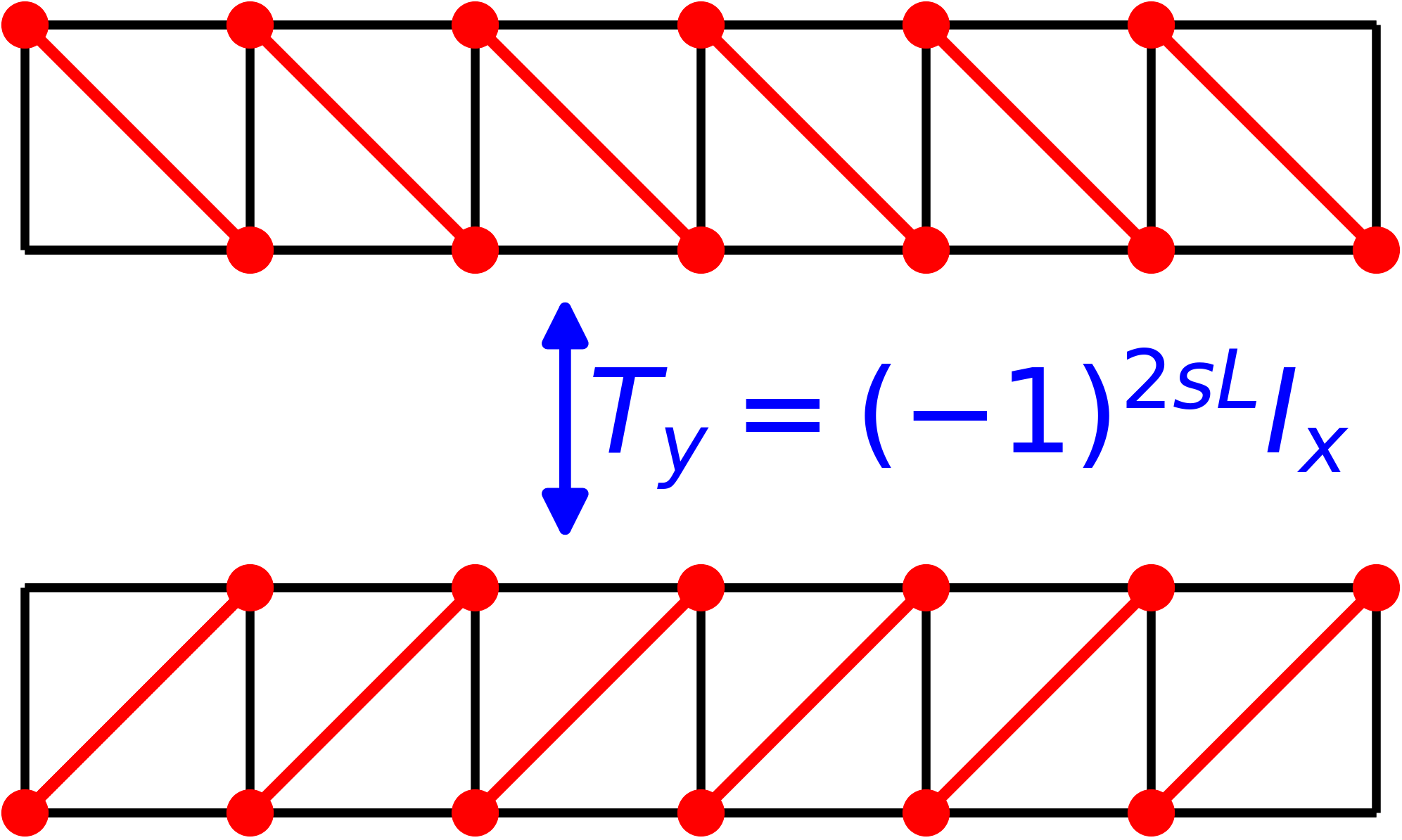}
  \caption{The exact valence-bond solid eigenstates of the $L \cross 2$ Heisenberg ladder and their relationship under symmetry. Red lines indicate valence bonds (a pair of lattice sites forming a spin singlet).}
  \label{fig:LadderVBS}
\end{figure}

$\ket{S_0}$ respects $x$-translation symmetry but breaks $y$-translation symmetry, so it has a symmetry partner
\begin{equation}
    \begin{aligned}
        \ket{S_1} =& T_y\ket{S_0}\\
        =&\bigotimes_{x=0}^{L-1} \ket{\Phi}_{\left(x, 1\right)\left(x+1,0\right)}.\\
    \end{aligned}
\end{equation}
$\ket{S}_0$ and $\ket{S}_1$ transform under $x$-inversion as
\begin{equation}
    \begin{aligned}
        I_x\ket{S_0}
        =&\bigotimes_{x=0}^{L-1} \ket{\Phi}_{\left(-x, 0\right)\left(-x-1,1\right)}\\
        =&(-1)^{2sL}\bigotimes_{x'=0}^{L-1} \ket{\Phi}_{\left(x',1\right)\left(x'+1, 0\right)}\\
        =&(-1)^{2sL}\ket{S_1}.\\
    \end{aligned}
\end{equation}
As tensor products of singlets, both $\ket{S_0}$ and $\ket{S_1}$ have zero total spin and spin-inversion eigenvalue $P_z\ket{S_{0,1}} = (-1)^{2sL}\ket{S_{0,1}}$.
Thus, the symmetry-sector eigenstates are the symmetric and antisymmetric linear combinations given Tab. \ref{tab:ladder_VBS}. For finite $L$, the two valence-bond solids are not orthogonal, so there is some normalization factor in front of the symmetric and antisymmetric linear combinations that we have omitted.

If one splits the full ladder into two subsystems by making two cuts (periodic boundary conditions) parallel to the $y$-axis, two valence bonds are cut, so the entanglement rank and entropy are $(2s+1)^2$ and $2 \log(2s+1)$ respectively. This is independent of the ladder's length $L$, so $\ket{S_0}$ and $\ket{S_1}$ have area-law entanglement.

In general, if $\ket{\psi_1}$ has entanglement rank $r_1$, and $\ket{\psi_2}$ has entanglement rank $r_2$, then $c_1\ket{\psi_1} + c_2\ket{\psi_2}$ has entanglement rank at most $r_1 + r_2$. Thus, the symmetry-sector eigenstates $\ket{S_\pm}$ also have area-law entanglement of at most $2\log(2s+1) + \log(2)$.

\begin{table}[h!]
    \centering
    \begin{tabular}{c|c|c|c|c|c|c}
    
        State & $k_x$ & $k_y$ & $I_x$ & $P_z$ & $J$ & $E$\\
        \hline
        \hline
        $\ket{S_+} = (1 + T_y) \ket{S_0}$ & $0$ & $0$ & $\chi$ & $\chi$ & $0$ & $0$ \\ 
        $\ket{S_-} = (1 - T_y) \ket{S_0}$ & $0$ & $\pi$ & $-\chi$ & $\chi$ & $0$ & $0$ \\ 
    \end{tabular}
    \caption{Exact valence-bond solid zero modes of the $L \cross 2$ Heisenberg ladder, where $\chi = (-1)^{2sL}$.}
    \label{tab:ladder_VBS}
\end{table}

\subsection{Even by Even Lattices}

We now consider fully two-dimensional systems of size $L_x \cross L_y$, where $L_x$ and $L_y$ are even and at least $4$.
We impose periodic boundary conditions $(x + L_x, y) \equiv (x, y)$ and $(x, y + L_y) \equiv (x, y)$.
The Hamiltonian can be factorized as
\begin{equation}
    \label{eq:square_Hamiltonian}
    \begin{aligned}
        H =&  \sum_{x=0}^{L_x - 1}\sum_{y=0}^{L_y - 1}
        \left( 
        \v S_{x,y} \cdot \v S_{x+1,y} + \v S_{x,y} \cdot \v S_{x,y + 1}
        \right)\\
        =&  
        \sum_{\substack{\text{Even}\\x=0}}^{L_x - 2}
        \sum_{\substack{\text{Even}\\y=0}}^{L_y - 2}
        \big[
        \v S_{(x+1,y)(x,y+1)} \cdot
        \v S_{(x,y)(x+1,y+1)} \\
        &+ 
        \v S_{(x+2,y+1)(x+1,y+2)} \cdot
        \v S_{(x+1,y+1)(x+2,y+2)}        
        \big].
    \end{aligned}
\end{equation}
An exact zero-energy eigenstate is the diagonal valence-bond solid
\begin{equation}
    \ket{S_{00}} = 
    \bigotimes_{\substack{\text{Even}\\x=0}}^{L_x - 2}
    \bigotimes_{\substack{\text{Even}\\y=0}}^{L_y - 2}
    \ket{\Phi}_{(x,y)(x+1,y+1)}
    \ket{\Phi}_{(x+2,y+1)(x+1,y+2)}.
\end{equation}
From the valence bonds, we have
\begin{equation}
    \begin{aligned}
        \v S_{(x,y)(x+1,y+1)} 
        \ket{S_{00}} &= 0,\\
        S_{(x+2,y+1)(x+1,y+2)}
        \ket{S_{00}} &= 0,\\
    \end{aligned}
\end{equation}
which combined with Eq. \ref{eq:square_Hamiltonian} proves that $H\ket{S_{00}}=0$.

$\ket{S_{00}}$ is only periodic relative to a $2 \cross 2$ supercell, so it is related to three other exact zero modes by symmetry
\begin{equation}
    \ket{S_{ab}} = T_x^a T_y^b \ket{S_{00}}, ab \in \{00, 01, 10, 11\}.
\end{equation}
$\ket{S_{00}}$ transforms under $x$-inversion as
\begin{equation}
    \begin{aligned}
        &
        I_x \ket{S_{00}}
        \\=&
        \bigotimes_{\substack{\text{Even}\\x=0}}^{L_x - 2}
        \bigotimes_{\substack{\text{Even}\\y=0}}^{L_y - 2}
        \ket{\Phi}_{(- x, y)(- x - 1, y + 1)}
        \ket{\Phi}_{(- x - 2, y + 1)(- x - 1, y + 2)}
        \\=&
        \bigotimes_{\substack{\text{Even}\\x'=0}}^{L_x - 2}
        \bigotimes_{\substack{\text{Even}\\y'=0}}^{L_y - 2}    
        \ket{\Phi}_{(x'+2,y'+2)(x'+1,y'+3)}
        \ket{\Phi}_{(x',y'+1)(x'+1,y'+2)}
        \\=& 
        T_y \ket{S_{00}}.
    \end{aligned}
\end{equation}
$\ket{S_{00}}$ transforms under $y$-inversion as
\begin{equation}
    \begin{aligned}
        &I_y \ket{S_{00}} 
        \\=&
        \bigotimes_{\substack{\text{Even}\\x=0}}^{L_x - 2}
        \bigotimes_{\substack{\text{Even}\\y=0}}^{L_y - 2}
        \ket{\Phi}_{(x, - y)(x + 1, - y - 1)}
        \ket{\Phi}_{(x + 2, - y - 1)(x + 1, - y - 2)}
        \\=&
        \bigotimes_{\substack{\text{Even} \\ x'=0}}^{L_x - 2}
        \bigotimes_{\substack{\text{Even} \\ y'=0}}^{L_y - 2}
        \ket{\Phi}_{(x'+2,y'+2)(x'+3,y'+1)}
        \ket{\Phi}_{(x'+2,y'+1)(x'+1,y')}
        \\=&
        T_x \ket{S_{00}}.
    \end{aligned}
\end{equation}
Thus, within the subspace spanned by the four valence-bond solids, $I_x = T_y$ and $I_y = T_x$. Because there are always an even number of valence bonds, all of the valence-bond solids have spin-inversion eigenvalue $P_z = 1 $ regardless of $s$.
Thus, the (unnormalized) symmetry-sector eigenstates are the linear combinations given in Tab. \ref{tab:even_even_VBS}.

\begin{figure}[t!]
  \centering
  \includegraphics[width=0.45\textwidth]{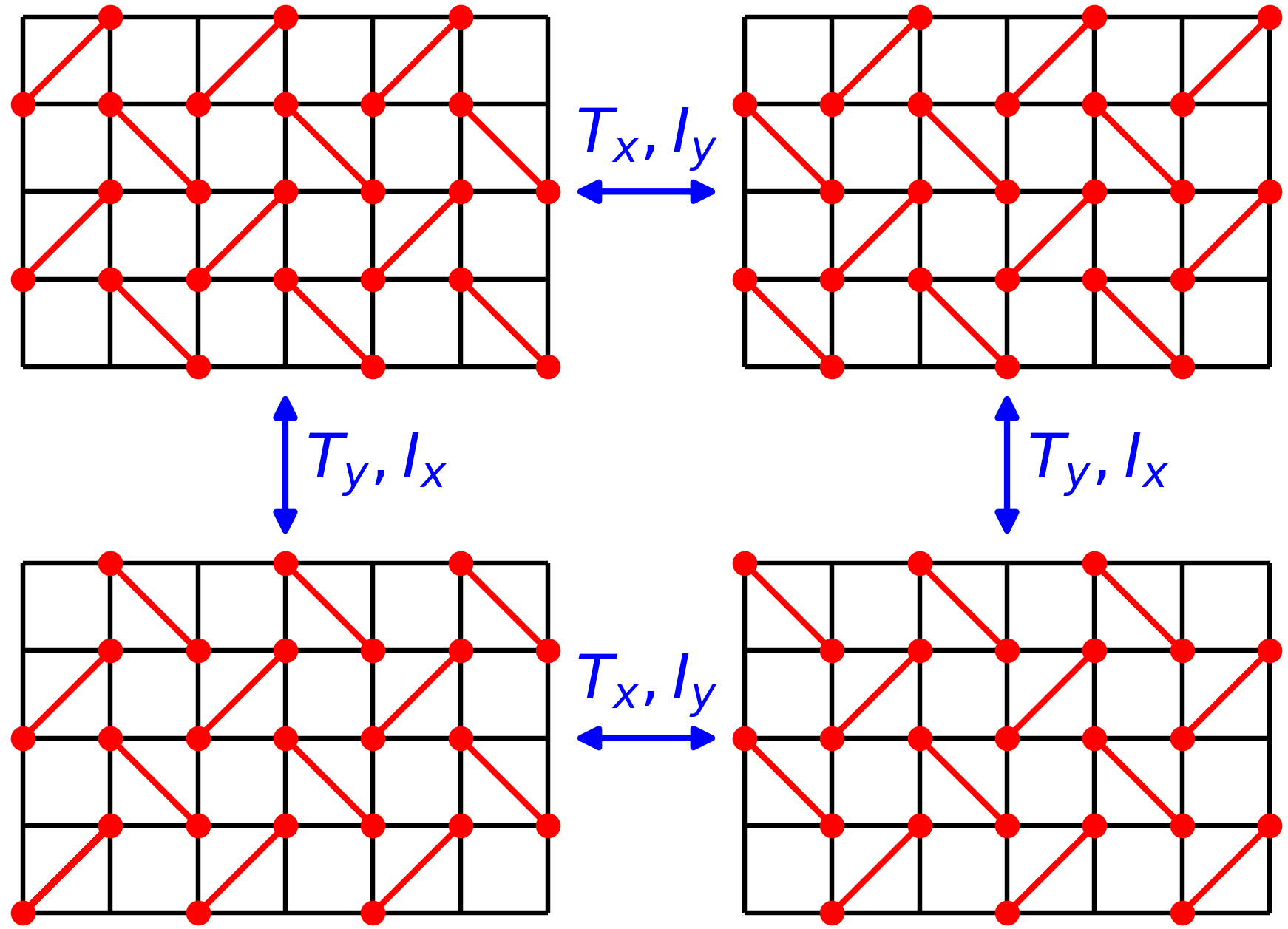}
  \caption{The exact valence-bond solid eigenstates of the even by even Heisenberg model and their relationship under symmetry. Red lines indicate valence bonds (a pair of lattice sites forming a spin singlet).}
  \label{fig:EvenEvenVBS}
\end{figure}

If one partitions the plane into two subsystems $A$ and $\Bar{A}$, the number of valence bonds cut is proportional to the size of the boundary $|\partial A|$, so our valence-bond solids $\ket{S_{ab}}$ have area-law entanglement. The same rank argument from before shows that the symmetry-sector eigenstates $\ket{S_{\pm\pm}}$ have entanglement entropy that is at most $\log(4)$ more than $\ket{S_{ab}}$ and thus also area law.

\begin{table}[ht!]
    \centering
    \begin{tabular}{c|c|c|c|c|c|c|c}
        State & $k_x$ & $k_y$ & $I_x$ & $I_y$ & $P_z$ & $J$ & $E$\\
        \hline
        \hline
        $\ket{S_{++}} = (1 + T_x ) (1 + T_y) \ket{S_{00}}$ & $0$ & $0$ & 1 & 1 & 1 & 0 & 0\\
        $\ket{S_{+-}} = (1 + T_x ) (1 - T_y) \ket{S_{00}}$ & 0 & $\pi$ & -1 & 1 & 1 & 0 & 0\\
        $\ket{S_{-+}} = (1 - T_x ) (1 + T_y) \ket{S_{00}}$ & $\pi$ & 0 & 1 & -1 & 1 & 0 & 0\\
        $\ket{S_{--}} = (1 - T_x ) (1 - T_y) \ket{S_{00}}$ & $\pi$ & $\pi$ & -1 & -1 & 1 & 0 & 0\\
    \end{tabular}
    \caption{Exact valence-bond solid zero modes in the square-lattice Heisenberg model.}
    \label{tab:even_even_VBS}
\end{table}

\subsection{Scars v.s. Hidden Symmetry}

Some systems host exact mid-spectrum eigenstates that are not scars, because they are instead the unique states in a hidden symmetry sector. For example, the Hubbard model on a bipartite lattice has states protected by an ``eta-pairing SU(2)'' symmetry \cite{Hubbard_SO4, extended_eta_scar}. Here, we rule out symmetry protection for the exact valence-bond solids.

From our exact diagonalization calculations, we find no other exact excited states except for states of a few magnons near the top (highest-energy states) of the spectrum. Specifically, the exact valence-bond solids are not part of a tower.
For the Heisenberg ladder, a hidden symmetry would have to mix between the states $\ket{S_\pm}$ and the ferromagnetic state $\ket{F}=\ket{s,s}^{\otimes 2L}$. In the spin-$1/2$ case, it is possible to write down a reasonably simple operator
\begin{equation}
    Q_S = \prod_{x=0}^{L-1} \left(S_{x, 0}^- - S_{x+1,1}^-\right).
\end{equation}
connecting $\ket{F}$ to $\ket{S_0}$.

However, $Q_S$ does not have a meaningful commutator with $H$. Furthermore, for higher spins, creating a valence bond $\ket{0,0}_{12}$ from $\ket{s,s}_{12}$ is more complicated. A similar situation exists for the even by even case. The presence of a hidden symmetry would also be inconsistent with the observed level statistics. Thus, the exact valence-bond solids are quantum many-body scars.

\section{Additional Scars in Ladders}

Remarkably, in even-length ladders $2l \cross 2$, two additional daughter exact states exist, corresponding to quasiparticles on top of a parent valence-bond solid.

\subsection{One-Magnon Daughter State}
This daughter state has $J=1$ and $E=0$.
It is a spin-$1$ magnon with wavevector $\pi$ on top of the parent valence-bond solid:
\begin{equation}
    \label{eq:ladder_magnon}
    \begin{aligned}
        Q_A &= \frac{1}{2} \sum_{x=0}^{L-1} (-1)^x \left( S^z_{x, 0} + S^z_{x, 1} \right)\\
        &= \frac{1}{2} \sum_{x=0}^{L-1} (-1)^x \left( S^z_{x, 0} - S^z_{x + 1, 1} \right),\\
        \ket{A_{0}} &= Q_A \ket{S_0}\\
        &= \sum_{x=0}^{L-1} (-1)^x S^z_{x, 0} \ket{S_0}.\\
    \end{aligned}
\end{equation}
We first show that $\left( S^z_{x, 0} - S^z_{x + 1, 1} \right)/2$ converts a valence bond $\ket{\Phi} = \ket{0,0}$ on sites $(x, 0)$ and $(x + 1, 1)$ into a magnon $\ket{1, 0}$. For brevity, we relabel $(x,0)\rightarrow1$ and $(x+1,1)\rightarrow2$. Also, note that $\left[\left( S^z_1 - S^z_2 \right)/2\right]\ket{\Phi}_{12} = S_1^z \ket{\Phi}_{12}$.
Then we have
\begin{equation}
    \begin{aligned}
        &S_{12}^2 S_1^z\ket{\Phi}_{12}\\
        =& \comm{S_{12}^2}{S_1^z}\ket{\Phi}_{12}\\
        =& \left( \v{S}_{12} \cdot \comm{\v{S}_{12}}{S_1^z} + 
        \comm{\v{S}_{12}}{S_1^z} \cdot \v S_{12} \right) \ket{\Phi}_{12}\\
        =& i\left( S_{12}^y \cdot S_1^x - S_{12}^x \cdot S_1^y  \right)\ket{\Phi}_{12}\\
        =& i\left( \comm{S_{12}^y}{S_1^x} - \comm{S_{12}^x}{S_1^y}  \right)\ket{\Phi}_{12}\\
        =&2 S_1^z\ket{\Phi}_{12},
    \end{aligned}
\end{equation}
proving that $S_1^z\ket{\Phi}_{12} \propto \ket{1,0}_{12}$.

\begin{figure}[t!]
  \centering
  \includegraphics[width=0.25\textwidth]{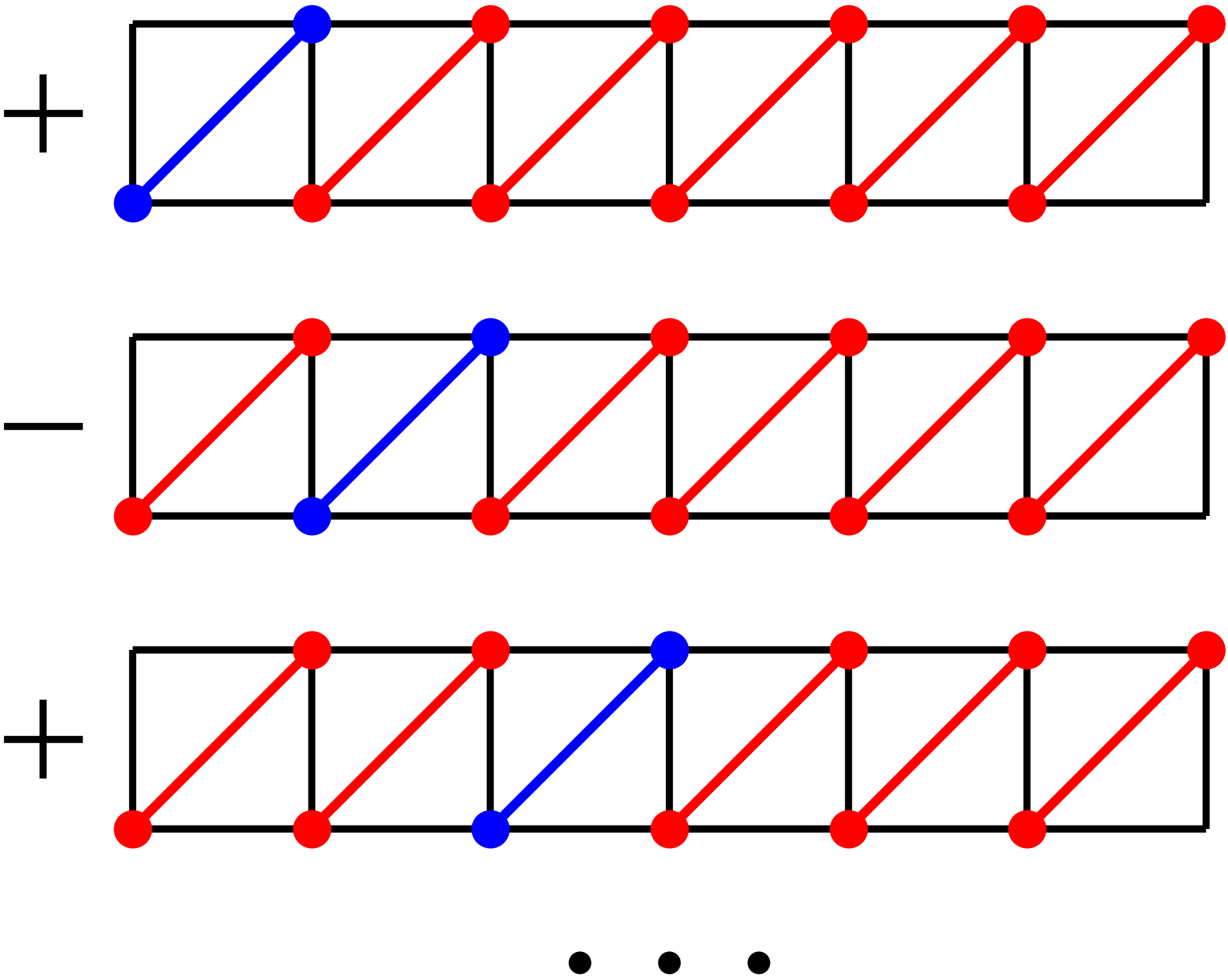}
  \caption{Exact valence-bond solid plus one-magnon eigenstate in the $L \cross 2$ Heisenberg ladder. Red lines indicate valence bonds (a pair of lattice sites forming a spin singlet), while a blue line indicates a pair of lattice forming a total spin-$1$ state.}
  \label{fig:MagnonVBS}
\end{figure}

$H$'s action on $\ket{A_0}$ is
\begin{equation}
    \label{eq:H_1Magnon}
    \begin{aligned}
        & H\ket{A_{0}}\\
        =& \sum_{x=0}^{L-1} \v{S}_{(x+1,0)(x,1)} \cdot \v S_{(x,0)(x+1,1)}  \sum_{x'=0}^{L-1} (-1)^{x'} S^z_{x', 0} \ket{S_0}\\
        =& \sum_{x=0}^{L-1} (-1)^x
        \big[\v{S}_{x+1,0} \cdot \v S_{(x,0)(x+1,1)} \, S^z_{x, 0}\\
        &- \v{S}_{x+1,1} \cdot \v S_{(x+1,0)(x+2,1)} \, S^z_{x+1, 0} \big] \ket{S_0}.\\
    \end{aligned} 
\end{equation}
We concentrate on the four sites present relevant to the above summand, labeling them $(x, 0) \rightarrow 1$, $(x+1,1) \rightarrow 2$, $(x+1,0) \rightarrow 3$, and $(x+2,1) \rightarrow 4$ for brevity. Then we have
\begin{equation}
    \begin{aligned}
        & \left[ \left( \v S_3 \cdot \v S_{12} \right) S_1^z - \left( \v S_2 \cdot \v S_{34} \right) S_3^z \right] \ket{\Phi}_{12} \ket{\Phi}_{34}\\
        =& \left[ \v S_3 \cdot \comm{\v S_{12}}{S_1^z} - \v S_2 \cdot \comm{\v S_{34}}{S_3^z} \right] \ket{\Phi}_{12} \ket{\Phi}_{34}\\
        =& i\left[ S_3^yS_1^x - S_3^xS_1^y - S_2^y S_3^x + S_2^x S_3^y \right] \ket{\Phi}_{12} \ket{\Phi}_{34}\\
        =&0,
    \end{aligned}
\end{equation}
proving that $H\ket{A_0} = 0$.

The magnon creation operator satisfies, $T_x Q_A T_x = - Q_A$, $T_y Q_A T_y = Q_A$, $P_z Q_A P_z = -Q_A$, and $I_x Q_A I_x= Q_A$. $Q_A$'s $T_y$-invariance implies that the symmetry-sector eigenstates are the symmetric and antisymmetric linear combinations given Tab. \ref{tab:ladder_one_magnon}. The above derivations only work when all operators act right on the valence-bond solid, so further applications of $Q_A$ do not generate additional exact excited states.

\begin{table}[ht!]
    \centering
    \begin{tabular}{c|c|c|c|c|c|c}
        State & $k_x$ & $k_y$ & $I_x$ & $P_z$ & $J$ & $E$\\
        \hline
        \hline
        $\ket{A_+} = (1 + T_y) Q_A \ket{S_0}$ & $\pi$ & $0$ & $1$ & $-1$ & $1$ & $0$ \\ 
        $\ket{A_-} = (1 - T_y) Q_A \ket{S_0}$ & $\pi$ & $\pi$ & $-1$ & $-1$ & $1$ & $0$ \\ 
    \end{tabular}
    \caption{Exact valence-bond solid plus one-magnon eigenstates in the $L \cross 2$ Heisenberg ladder. Because $L$ is even, the previously defined $\chi = (-1)^{2sL}=1$.}
    \label{tab:ladder_one_magnon}
\end{table}

\subsection{Two-Magnon Daughter State}
Although a full tower of magnons does not exist, there is an additional two-magnon exact eigenstate with $J=0$ and $E=-2$. It has two spin-$1$ magnons which together form a spin-singlet bound state on top of the valence-bond solid:
\begin{equation}
    \label{eq:ladder_two_magnon}
    \begin{aligned}
        Q_B &= \sum_{x=0}^{L-1} (-1)^x \left[ \v S_{x,0} \cdot \v S_{x,1} \right],\\
        \ket{B_0} &= Q_B \ket{S_0}.\\
    \end{aligned}
\end{equation}
We concentrate on the four sites relevant to the above summand, labeling them $(x-1, 0) \rightarrow 1$, $(x,1) \rightarrow 2$, $(x,0) \rightarrow 3$, and $(x+1,1) \rightarrow 4$ for brevity. The state on $1234$ is $\v S_2 \cdot \v S_3 \ket{\Phi}_{12} \ket{\Phi}_{34}$. We proved earlier that $S_1^z\ket{\Phi}_{12} = -S_2^z\ket{\Phi}_{12} \propto \ket{J=1,M=0}_{12}$, and likewise $S_1^\pm\ket{\Phi}_{12} = -S_2^\pm\ket{\Phi}_{12} \propto \ket{J=1,M=\pm1}_{12}$. Thus, $\v S_2 \cdot \v S_3$ creates two spin-$1$ magnons on sites $12$ and $34$. Because $\v S_2 \cdot \v S_3$ is rotationally invariant, the two magnons together must form a singlet.

$H$'s action on $\ket{B_0}$ is:
\begin{equation}
    \label{eq:H_on_magnon_pair}
    \begin{aligned}
        H\ket{B_{0}} =& \sum_{x=0}^{L-1} \sum_{x'=0}^{L-1} (-1)^{x'} \big[\v{S}_{(x+1,0)(x,1)} \cdot \v S_{(x,0)(x+1,1)} \\
        & \v S_{x',0} \cdot \v S_{x',1}\ket{S_0}\big]\\
        =& \sum_{x=0}^{L-1} (-1)^x \big[\v{S}_{(x+1,0)(x,1)} \cdot \v S_{(x,0)(x+1,1)}\\
        &\left(\v S_{x,0} \cdot \v S_{x,1} - \v S_{x+1,0} \cdot \v S_{x+1,1} \right) \big] \ket{S_0}.\\
    \end{aligned} 
\end{equation}
We concentrate on the six sites relevant to the above summand, labeling them $(x-1,0) \rightarrow 1$, $(x,1) \rightarrow 2$, $(x,0) \rightarrow 3$, $(x+1,1) \rightarrow 4$, $(x+1,0) \rightarrow 5$, and $(x+2,1) \rightarrow 6$. Then we have:
\begin{equation}
    \begin{aligned}
        &\left( \v S_{25} \cdot \v S_{34} \right) \left( \v S_2 \cdot \v S_3 - \v S_4 \cdot \v S_5 \right) \ket{\Phi}_{12}\ket{\Phi}_{34}\ket{\Phi}_{56}\\
        =&\left( \v S_{25} \cdot \v S_{34} \right) \left( \v S_{25} \cdot \v S_3 \right) \ket{\Phi}_{12}\ket{\Phi}_{34}\ket{\Phi}_{56}\\
        =&S_{25}^\mu S_{34}^\mu S_{25}^\nu S_3^\nu \ket{\Phi}_{12}\ket{\Phi}_{34}\ket{\Phi}_{56}\\
        =&S_{25}^\mu S_{25}^\nu \comm{S_{34}^\mu}{S_3^\nu} \ket{\Phi}_{12}\ket{\Phi}_{34}\ket{\Phi}_{56}\\
        =&i\epsilon^{\mu\nu\kappa}S_{25}^\mu S_{25}^\nu S_3^\kappa \ket{\Phi}_{12}\ket{\Phi}_{34}\ket{\Phi}_{56}\\
        =&-S_{25}^\kappa S_3^\kappa \ket{\Phi}_{12}\ket{\Phi}_{34}\ket{\Phi}_{56}\\
        =&-\left( \v S_2 \cdot \v S_3 - \v S_4 \cdot \v S_5 \right) \ket{\Phi}_{12}\ket{\Phi}_{34}\ket{\Phi}_{56}.\\
    \end{aligned}
\end{equation}
Summing over all $x$ yields $H\ket{B_0} = -2\ket{B_0}$. The negative energy reflects the fact that the magnons are anti-aligned to form a singlet. 

\begin{figure}[t!]
  \centering
  \includegraphics[width=0.25\textwidth]{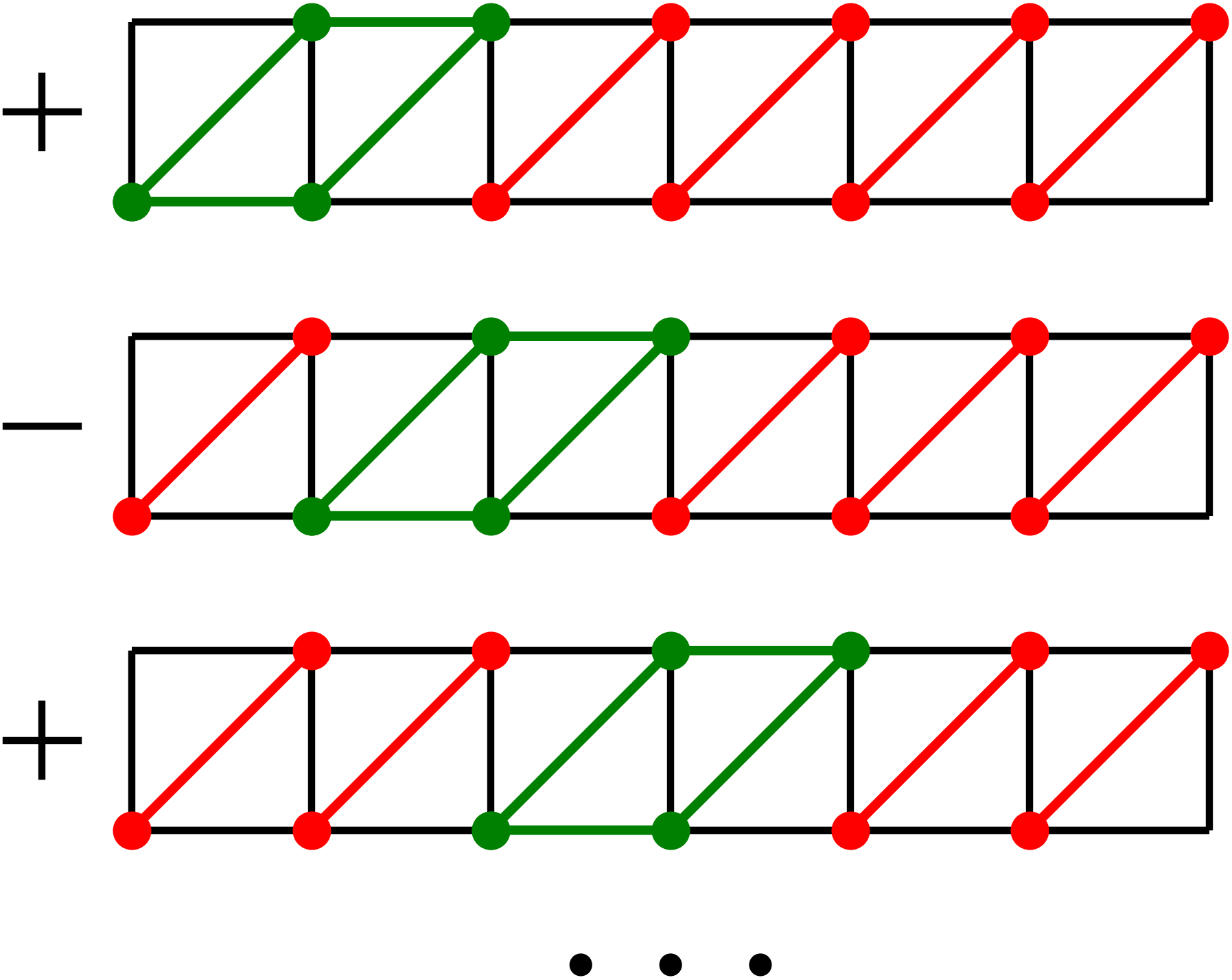}
  \caption{Exact valence-bond solid plus two-magnon eigenstate in the $L \cross 2$ Heisenberg ladder. Red lines indicate valence bonds (a pair of lattice sites forming a spin singlet), while a green parallelogram indicates two spin-$1$ states on the diagonals adding into a total spin-$0$ state.}
  \label{fig:TwoMagnonVBS}
\end{figure}

\begin{table}[ht!]
    \centering
    \begin{tabular}{c|c|c|c|c|c|c}
        State & $k_x$ & $k_y$ & $I_x$ & $P_z$ & $J$ & $E$\\
        \hline
        \hline
        $\ket{B_+} = (1 + T_y) Q_B \ket{S_0}$ & $\pi$ & $0$ & $1$ & $1$ & $0$ & $-2$ \\ 
        $\ket{B_-} = (1 - T_y) Q_B \ket{S_0}$ & $\pi$ & $\pi$ & $-1$ & $1$ & $0$ & $-2$ \\ 
    \end{tabular}
    \caption{Exact valence-bond solid plus two-magnon zero modes in the $L \cross 2$ Heisenberg ladder. Because $L$ is even, the previously defined $\chi = (-1)^{2sL}=1$.}
    \label{tab:ladder_two_magnon}
\end{table}

The magnon-pair creation operator satisfies, $T_x Q_B T_x = - Q_B$, $T_y Q_B T_y = Q_B$, $P_z Q_B P_z = Q_B$, and $I_x Q_B I_x= Q_B$. Invariance under $T_y$ implies that $\ket{B_+} = Q_B \ket{S_+}$ and $\ket{B_-} = Q_B \ket{S_-}$ are also exact eigenstates in sectors listed in Tab. \ref{tab:ladder_two_magnon}.

\section{Connection to Projector Embedding}

We note that the mechanism for scarring in our system is reminiscent of the Shiraishi-Mori projector embedding \cite{ShiraishiMori}. In their case, the Hamiltonian is
\begin{equation}
    \label{eq:projector_embedding}
    H = \sum_i P_i h_i P_i + H'.
\end{equation}
$P_i$ are local projectors that do not necessarily commute, but for which there exists a nontrivial common nullspace $P_i\mathcal{T}=0$. The $h_i$ are arbitrary, while $\comm{P_i}{H'} = 0$.

The eigenstates of $H'$ within the subspace $\mathcal{T}$ are the many-body scars. They can be placed in the middle of the spectrum by varying $H'$ and $h_i$; for a generic choice of $P_i$, $h_i$, and $H'$, Eq. \ref{eq:projector_embedding} will be nonintegrable. The scars are anomalous because they have exactly zero $P_i$ expectation value without fluctuation.

\begin{figure}[t!]
  \centering
  \includegraphics[width=0.45\textwidth]{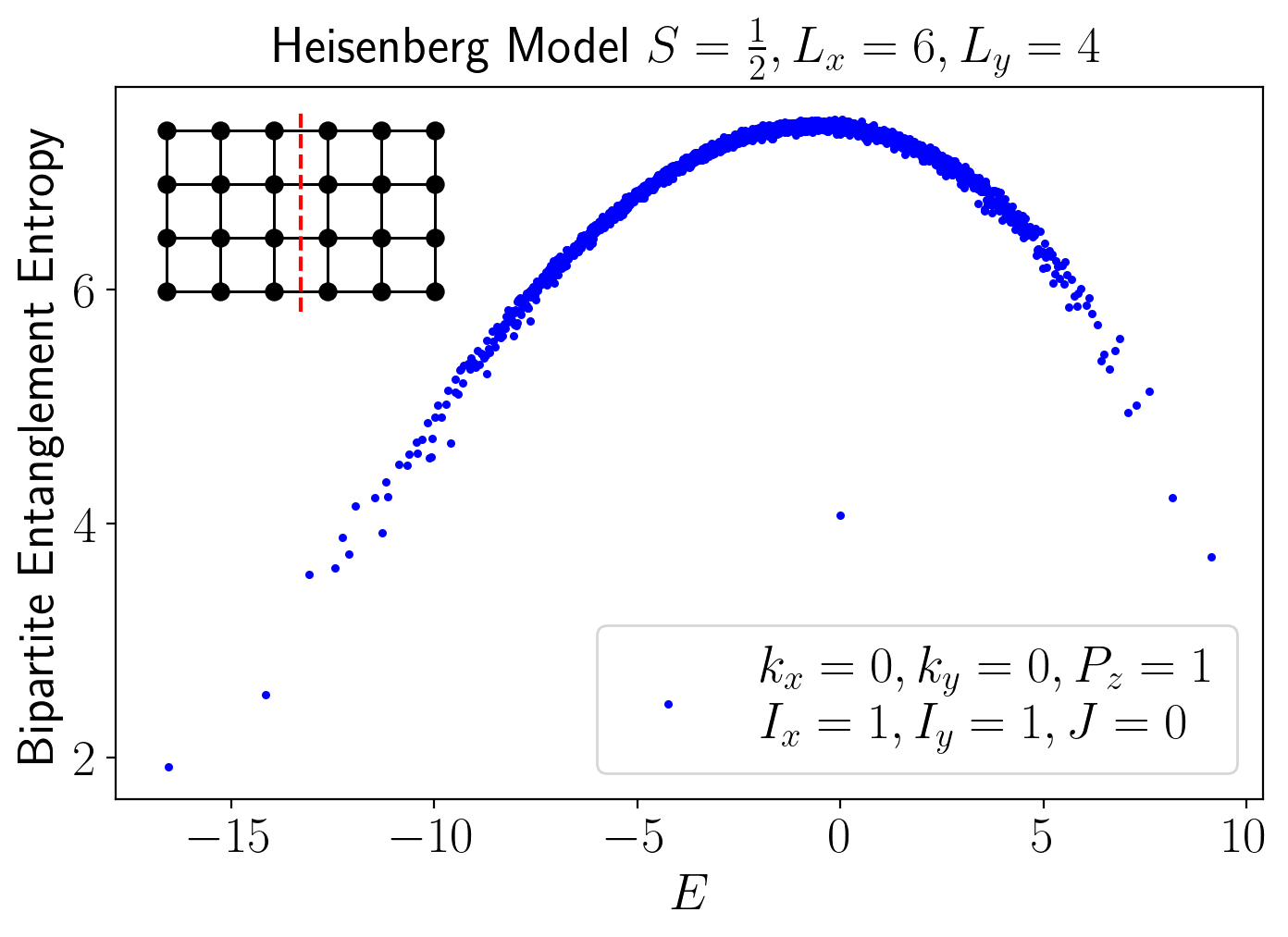}
  \caption{Bipartite entanglement entropy of the valence-bond solid scar $\ket{S_{++}}$ and all other eigenstates in its quantum number sector for the $S=\frac{1}{2}, L_x = 6, L_y = 4$ Heisenberg model.}
  \label{fig:64_S++_entanglement}
\end{figure}

For the ladder, the Hamiltonian can be placed into the Shiraishi-Mori form
\begin{equation}
    \label{eq:ladder_projector}
    \begin{aligned}
        P_x &= 
        \left(1 - P_{(x + 1,0)(x,1)}^{J = 0}\right) \left(1 - P_{(x,0)(x+1,1)}^{J = 0}\right),\\
        H &= \sum_{x=0}^{L-1} P_x\v{S}_{(x+1,0)(x,1)} \cdot \v S_{(x,0)(x+1,1)} P_x,
    \end{aligned}
\end{equation}
where $P^{J = 0}_{ij} = \ketbra{\Phi}{\Phi}_{ij}$ is the projector onto the spin-singlet subspace on sites $i$ and $j$. The two valence-bond solids $\ket{S_{0,1}}$ are the only states annihilated by all $P_x$. To be annihilated by $P_x$, $\ket{\Psi}$ must have a valence bond either on $(x,0)(x+1,1)$ or $(x + 1,0)(x,1)$. Once we choose which valence bond to place on plaquette $x$, plaquettes $x-1$ and $x+1$ are fixed, and so forth. It is remarkable that Eq. \ref{eq:ladder_projector}, which was originally constructed specifically to violate the eigenstate thermalization hypothesis \cite{ShiraishiMori}, also can describe a realistic many-body Hamiltonian. However, the Eq. \ref{eq:ladder_projector} does not capture the four daughter quasiparticle states $\ket{A_{0,1}}$ and $\ket{B_{0,1}}$.

For the even-by-even square case, based on Eq. \ref{eq:square_Hamiltonian}, we could write the Shiraishi-Mori form
\begin{equation}
    \begin{aligned}
        P_{x,y} = & 
        \left(1 - 
        P_{(x+2,y+1)(x+1,y+2)}^{J = 0}
        P_{(x,y)(x+1,y+1)}^{J = 0}
        \right)\\
        &
        \left(1 - 
        P_{(x+1,y+1)(x+2,y+2)}^{J = 0}
        P_{(x+1,y)(x,y+1)}^{J = 0}
        \right),
        \\
        H =& 
        \sum_{\substack{\text{Even}\\x=0}}^{L_x - 2}
        \sum_{\substack{\text{Even}\\y=0}}^{L_y - 2}
        \big[P_{x,y}
        \big(
        \v S_{(x+1,y)(x,y+1)} \cdot
        \v S_{(x,y)(x+1,y+1)} \\
        &+ 
        \v S_{(x+2,y+1)(x+1,y+2)} \cdot
        \v S_{(x+1,y+1)(x+2,y+2)}\big)
        P_{x,y}\big].
    \end{aligned}
\end{equation}
However, while $\ket{S_{00}}$ and $\ket{S_{11}}$ lie in the common nullspace of $\{P_{x,y} : \text{even } x \text{ and } y\}$, $\ket{S_{01}}$ and $\ket{S_{10}}$ do not. This is because $\ket{S_{00}}$ and $\ket{S_{11}}$ have valence bonds on even plaquettes (those whose lower left corner lies on the even sublattice), while $\ket{S_{01}}$ and $\ket{S_{10}}$ have valence bonds on odd plaquettes.

\begin{figure}[t!]
  \centering
  \includegraphics[width=0.45\textwidth]{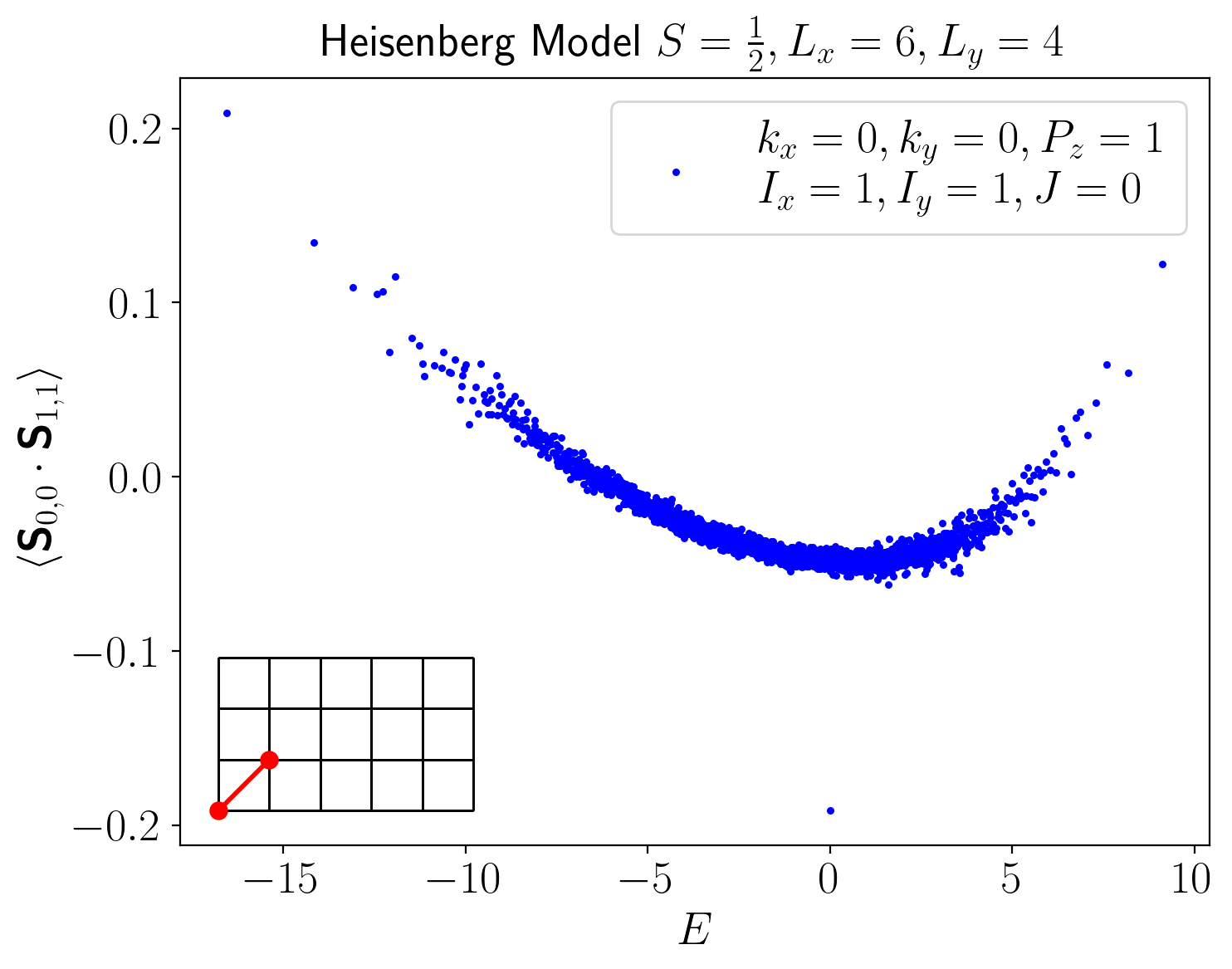}
  \caption{Next-nearest-neighbor correlation function of the valence-bond solid scar $\ket{S_{++}}$ and all other eigenstates in its quantum number sector for the $S=\frac{1}{2}, L_x = 6, L_y = 4$ Heisenberg model.}
  \label{fig:64_S++_correlator}
\end{figure}

To correct this, one could try the projectors
\begin{equation}
    \label{eq:22_projectors}
    \begin{aligned}
        P'_{x,y} =
        &\left(1 - 
        P_{(x+2,y+1)(x+1,y+2)}^{J = 0}
        P_{(x,y)(x+1,y+1)}^{J = 0}
        \right)\\
        &\left(1 - 
        P_{(x+1,y+1)(x+2,y+2)}^{J = 0}
        P_{(x+1,y)(x,y+1)}^{J = 0}
        \right)\\
        &\left(1 - 
        P_{(x+1,y+1)(x,y+2)}^{J = 0}
        P_{(x+1,y)(x+2,y+1)}^{J = 0}
        \right)\\
        &\left(1 - 
        P_{(x,y+1)(x+1,y+2)}^{J = 0}
        P_{(x+2,y)(x+1,y+1)}^{J = 0}
        \right).\\
    \end{aligned}
\end{equation}
The nullspace of $\{P_{x,y}' : \text{all } x \text{ and } y\}$ is spanned by the four possible $2 \cross 2$ supercells of the valence-bond solid, and the common nullspace of all $P'_{x,y}$ is $\text{span}\{\ket{S_{00}}, \ket{S_{10}}, \ket{S_{01}}, \ket{S_{11}}\}$ from stitching all $2 \cross 2$ patches together. However, because each $P'_{x,y}$ acts on a $2 \cross 2$ supercell, and each of the four factors in Eq. \ref{eq:22_projectors} corresponds to the cancellation of different pairs of terms in the Hamiltonian when acting on our valence-bond solids, it is not clear what $h_i$ would recover the Heisenberg Hamiltonian.

\section{Exact Diagonalization Results}

We performed comprehensive exact diagonalization calculations to confirm our analytical results and determine if our valence-bond solids are the only scars in the square-lattice Heisenberg model. The predicted exact eigenstates are observed in the correct symmetry sectors, and no other low-spin exact eigenstates are found.

For spin-$\frac{1}{2}$, we studied ladders of length $L=7-12$. For ladders, besides $L=8$, the only exact states had total spin $J \geq L-3$, corresponding to at most three magnons on top of the ferromagnetic state. For $L=8$, there were three eigenstates with $J=3$ and $E=0$. Because we observe no corresponding states in larger $L$, we believe that the $J=3$, $E=0$ states observed at $L=8$ are finite-size effects. Thus, our valence-bond solids cover all exact states of finite energy density (relative to either the bottom or top of the spectrum) in Heisenberg ladders.

For spin-$\frac{1}{2}$, we studied a $6$ by $4$ system. Aside from the four valence-bond solids predicted, all other exact excited had total spin $J \geq 8$, corresponding to at most four magnons excited on top of the ferromagnetic state. As for the ladder, the lack of low-spin exact excited states provides strong evidence that our valence-bond solids cover all exact states of finite energy density in Heisenberg.

For spin-$1$ systems, we were only able to study ladders up to $L = 7$ due to the larger site dimension. Nevertheless, the conclusions are the same as for spin-$1/2$: the only other exact eigenstates have total spin $J \geq 2L - 3$, corresponding to three magnons relative to the ferromagnetic state.

\section{Conclusion}

Our work shows that many-body scars exist in one of the most celebrated quantum antiferromagnets. To the best of our knowledge, our work is the first to discover nontrivial exact excited states in the square-lattice Heisenberg model. To the best of our knowledge, our work is also the first to discover many-body scars in a family of models having different site spin but otherwise identical Hamiltonians \footnote{We note that the AKLT chain also has many-body scars for higher spins \cite{AKLT_unified}, but the form of the AKLT interaction (i.e. expansion as a polynomial of $\v S_i \cdot \v S_{i+1}$) depends on the site spin.}.

Our work raises interesting questions about the connection between quantum many-body scars in lattice models and the semiclassical origin of one-particle scars \cite{scar_prl, scar_smoothed, scar_linearity, scar_review}; better understanding this connection (or possibly the lack thereof) has been a longstanding goal within the scar community \cite{scars_qft}. The fact that our scars exist for Heisenberg models of all spins allows us to take the large-$S$ limit. In the context of ground-state physics, this is generally regarded as making the making the model more semiclassical. For example, one may construct spin coherent states \cite{spin_coherent, spin_coherent_2}, which become increasingly localized as $S$ increases. Nevertheless, because our valence-bond solids are composed of maximally-entangled singlets, the scar state is firmly quantum-mechanical even in the large-$S$ limit. Still, it could be interesting to apply tools such as the spin coherent-state path integral to the dynamics of quantum antiferromagnets. 

We hope that our work stimulates interest in excited states and nonequilibrium dynamics within the condensed matter theory community, and interest in celebrated condensed matter systems such as the 2D Heisenberg models within the quantum dynamics and quantum information community. 

\section*{Acknowledgments} We thank Aram Harrow, Liang Fu, and Soonwon Choi for helpful discussions. DDD was supported by the Undergraduate Research Opportunities Program at MIT. The authors acknowledge the MIT SuperCloud and Lincoln Laboratory Supercomputing Center for providing high-performance computing resources.

\bibliography{references.bib}

\begin{thebibliography}{27}
\expandafter\ifx\csname natexlab\endcsname\relax\def\natexlab#1{#1}\fi
\expandafter\ifx\csname bibnamefont\endcsname\relax
  \def\bibnamefont#1{#1}\fi
\expandafter\ifx\csname bibfnamefont\endcsname\relax
  \def\bibfnamefont#1{#1}\fi
\expandafter\ifx\csname citenamefont\endcsname\relax
  \def\citenamefont#1{#1}\fi
\expandafter\ifx\csname url\endcsname\relax
  \def\url#1{\texttt{#1}}\fi
\expandafter\ifx\csname urlprefix\endcsname\relax\def\urlprefix{URL }\fi
\providecommand{\bibinfo}[2]{#2}
\providecommand{\eprint}[2][]{\url{#2}}

\bibitem[{\citenamefont{Bernien et~al.}(2017)\citenamefont{Bernien, Schwartz, Keesling, Levine, Omran, Pichler, Choi, Zibrov, Endres, Greiner et~al.}}]{scar_first_exp}
\bibinfo{author}{\bibfnamefont{H.}~\bibnamefont{Bernien}}, \bibinfo{author}{\bibfnamefont{S.}~\bibnamefont{Schwartz}}, \bibinfo{author}{\bibfnamefont{A.}~\bibnamefont{Keesling}}, \bibinfo{author}{\bibfnamefont{H.}~\bibnamefont{Levine}}, \bibinfo{author}{\bibfnamefont{A.}~\bibnamefont{Omran}}, \bibinfo{author}{\bibfnamefont{H.}~\bibnamefont{Pichler}}, \bibinfo{author}{\bibfnamefont{S.}~\bibnamefont{Choi}}, \bibinfo{author}{\bibfnamefont{A.~S.} \bibnamefont{Zibrov}}, \bibinfo{author}{\bibfnamefont{M.}~\bibnamefont{Endres}}, \bibinfo{author}{\bibfnamefont{M.}~\bibnamefont{Greiner}}, \bibnamefont{et~al.}, \bibinfo{journal}{Nature} \textbf{\bibinfo{volume}{551}}, \bibinfo{pages}{579–584} (\bibinfo{year}{2017}).

\bibitem[{\citenamefont{Turner et~al.}(2018{\natexlab{a}})\citenamefont{Turner, Michailidis, Abanin, Serbyn, and Papić}}]{PXP_theory_1}
\bibinfo{author}{\bibfnamefont{C.~J.} \bibnamefont{Turner}}, \bibinfo{author}{\bibfnamefont{A.~A.} \bibnamefont{Michailidis}}, \bibinfo{author}{\bibfnamefont{D.~A.} \bibnamefont{Abanin}}, \bibinfo{author}{\bibfnamefont{M.}~\bibnamefont{Serbyn}}, \bibnamefont{and} \bibinfo{author}{\bibfnamefont{Z.}~\bibnamefont{Papić}}, \bibinfo{journal}{Nature Physics} \textbf{\bibinfo{volume}{14}}, \bibinfo{pages}{745–749} (\bibinfo{year}{2018}{\natexlab{a}}).

\bibitem[{\citenamefont{Turner et~al.}(2018{\natexlab{b}})\citenamefont{Turner, Michailidis, Abanin, Serbyn, and Papi\ifmmode~\acute{c}\else \'{c}\fi{}}}]{PXP_theory_2}
\bibinfo{author}{\bibfnamefont{C.~J.} \bibnamefont{Turner}}, \bibinfo{author}{\bibfnamefont{A.~A.} \bibnamefont{Michailidis}}, \bibinfo{author}{\bibfnamefont{D.~A.} \bibnamefont{Abanin}}, \bibinfo{author}{\bibfnamefont{M.}~\bibnamefont{Serbyn}}, \bibnamefont{and} \bibinfo{author}{\bibfnamefont{Z.}~\bibnamefont{Papi\ifmmode~\acute{c}\else \'{c}\fi{}}}, \bibinfo{journal}{Phys. Rev. B} \textbf{\bibinfo{volume}{98}}, \bibinfo{pages}{155134} (\bibinfo{year}{2018}{\natexlab{b}}), \urlprefix\url{https://link.aps.org/doi/10.1103/PhysRevB.98.155134}.

\bibitem[{\citenamefont{Srednicki}(1994)}]{ETH1}
\bibinfo{author}{\bibfnamefont{M.}~\bibnamefont{Srednicki}}, \bibinfo{journal}{Phys. Rev. E} \textbf{\bibinfo{volume}{50}}, \bibinfo{pages}{888} (\bibinfo{year}{1994}), \urlprefix\url{https://link.aps.org/doi/10.1103/PhysRevE.50.888}.

\bibitem[{\citenamefont{Deutsch}(1991)}]{ETH2}
\bibinfo{author}{\bibfnamefont{J.~M.} \bibnamefont{Deutsch}}, \bibinfo{journal}{Phys. Rev. A} \textbf{\bibinfo{volume}{43}}, \bibinfo{pages}{2046} (\bibinfo{year}{1991}), \urlprefix\url{https://link.aps.org/doi/10.1103/PhysRevA.43.2046}.

\bibitem[{\citenamefont{Rigol et~al.}(2008)\citenamefont{Rigol, Dunjko, and Olshanii}}]{ETH3}
\bibinfo{author}{\bibfnamefont{M.}~\bibnamefont{Rigol}}, \bibinfo{author}{\bibfnamefont{V.}~\bibnamefont{Dunjko}}, \bibnamefont{and} \bibinfo{author}{\bibfnamefont{M.}~\bibnamefont{Olshanii}}, \bibinfo{journal}{Nature} \textbf{\bibinfo{volume}{452}}, \bibinfo{pages}{854–858} (\bibinfo{year}{2008}).

\bibitem[{\citenamefont{Lin and Motrunich}(2019)}]{PXP_exact}
\bibinfo{author}{\bibfnamefont{C.-J.} \bibnamefont{Lin}} \bibnamefont{and} \bibinfo{author}{\bibfnamefont{O.~I.} \bibnamefont{Motrunich}}, \bibinfo{journal}{Phys. Rev. Lett.} \textbf{\bibinfo{volume}{122}}, \bibinfo{pages}{173401} (\bibinfo{year}{2019}), \urlprefix\url{https://link.aps.org/doi/10.1103/PhysRevLett.122.173401}.

\bibitem[{\citenamefont{Choi et~al.}(2019)\citenamefont{Choi, Turner, Pichler, Ho, Michailidis, Papi\ifmmode~\acute{c}\else \'{c}\fi{}, Serbyn, Lukin, and Abanin}}]{PXP_su2}
\bibinfo{author}{\bibfnamefont{S.}~\bibnamefont{Choi}}, \bibinfo{author}{\bibfnamefont{C.~J.} \bibnamefont{Turner}}, \bibinfo{author}{\bibfnamefont{H.}~\bibnamefont{Pichler}}, \bibinfo{author}{\bibfnamefont{W.~W.} \bibnamefont{Ho}}, \bibinfo{author}{\bibfnamefont{A.~A.} \bibnamefont{Michailidis}}, \bibinfo{author}{\bibfnamefont{Z.}~\bibnamefont{Papi\ifmmode~\acute{c}\else \'{c}\fi{}}}, \bibinfo{author}{\bibfnamefont{M.}~\bibnamefont{Serbyn}}, \bibinfo{author}{\bibfnamefont{M.~D.} \bibnamefont{Lukin}}, \bibnamefont{and} \bibinfo{author}{\bibfnamefont{D.~A.} \bibnamefont{Abanin}}, \bibinfo{journal}{Phys. Rev. Lett.} \textbf{\bibinfo{volume}{122}}, \bibinfo{pages}{220603} (\bibinfo{year}{2019}), \urlprefix\url{https://link.aps.org/doi/10.1103/PhysRevLett.122.220603}.

\bibitem[{\citenamefont{Ho et~al.}(2019)\citenamefont{Ho, Choi, Pichler, and Lukin}}]{PXP_orbit}
\bibinfo{author}{\bibfnamefont{W.~W.} \bibnamefont{Ho}}, \bibinfo{author}{\bibfnamefont{S.}~\bibnamefont{Choi}}, \bibinfo{author}{\bibfnamefont{H.}~\bibnamefont{Pichler}}, \bibnamefont{and} \bibinfo{author}{\bibfnamefont{M.~D.} \bibnamefont{Lukin}}, \bibinfo{journal}{Phys. Rev. Lett.} \textbf{\bibinfo{volume}{122}}, \bibinfo{pages}{040603} (\bibinfo{year}{2019}), \urlprefix\url{https://link.aps.org/doi/10.1103/PhysRevLett.122.040603}.

\bibitem[{\citenamefont{Moudgalya et~al.}(2018{\natexlab{a}})\citenamefont{Moudgalya, Rachel, Bernevig, and Regnault}}]{AKLT_exact}
\bibinfo{author}{\bibfnamefont{S.}~\bibnamefont{Moudgalya}}, \bibinfo{author}{\bibfnamefont{S.}~\bibnamefont{Rachel}}, \bibinfo{author}{\bibfnamefont{B.~A.} \bibnamefont{Bernevig}}, \bibnamefont{and} \bibinfo{author}{\bibfnamefont{N.}~\bibnamefont{Regnault}}, \bibinfo{journal}{Phys. Rev. B} \textbf{\bibinfo{volume}{98}}, \bibinfo{pages}{235155} (\bibinfo{year}{2018}{\natexlab{a}}), \urlprefix\url{https://link.aps.org/doi/10.1103/PhysRevB.98.235155}.

\bibitem[{\citenamefont{Moudgalya et~al.}(2018{\natexlab{b}})\citenamefont{Moudgalya, Regnault, and Bernevig}}]{AKLT_entanglement}
\bibinfo{author}{\bibfnamefont{S.}~\bibnamefont{Moudgalya}}, \bibinfo{author}{\bibfnamefont{N.}~\bibnamefont{Regnault}}, \bibnamefont{and} \bibinfo{author}{\bibfnamefont{B.~A.} \bibnamefont{Bernevig}}, \bibinfo{journal}{Phys. Rev. B} \textbf{\bibinfo{volume}{98}}, \bibinfo{pages}{235156} (\bibinfo{year}{2018}{\natexlab{b}}), \urlprefix\url{https://link.aps.org/doi/10.1103/PhysRevB.98.235156}.

\bibitem[{\citenamefont{Mark et~al.}(2020)\citenamefont{Mark, Lin, and Motrunich}}]{AKLT_unified}
\bibinfo{author}{\bibfnamefont{D.~K.} \bibnamefont{Mark}}, \bibinfo{author}{\bibfnamefont{C.-J.} \bibnamefont{Lin}}, \bibnamefont{and} \bibinfo{author}{\bibfnamefont{O.~I.} \bibnamefont{Motrunich}}, \bibinfo{journal}{Phys. Rev. B} \textbf{\bibinfo{volume}{101}}, \bibinfo{pages}{195131} (\bibinfo{year}{2020}), \urlprefix\url{https://link.aps.org/doi/10.1103/PhysRevB.101.195131}.

\bibitem[{\citenamefont{Moudgalya et~al.}(2020)\citenamefont{Moudgalya, O'Brien, Bernevig, Fendley, and Regnault}}]{MPS_scar}
\bibinfo{author}{\bibfnamefont{S.}~\bibnamefont{Moudgalya}}, \bibinfo{author}{\bibfnamefont{E.}~\bibnamefont{O'Brien}}, \bibinfo{author}{\bibfnamefont{B.~A.} \bibnamefont{Bernevig}}, \bibinfo{author}{\bibfnamefont{P.}~\bibnamefont{Fendley}}, \bibnamefont{and} \bibinfo{author}{\bibfnamefont{N.}~\bibnamefont{Regnault}}, \bibinfo{journal}{Phys. Rev. B} \textbf{\bibinfo{volume}{102}}, \bibinfo{pages}{085120} (\bibinfo{year}{2020}), \urlprefix\url{https://link.aps.org/doi/10.1103/PhysRevB.102.085120}.

\bibitem[{\citenamefont{Schecter and Iadecola}(2019)}]{XY_1st}
\bibinfo{author}{\bibfnamefont{M.}~\bibnamefont{Schecter}} \bibnamefont{and} \bibinfo{author}{\bibfnamefont{T.}~\bibnamefont{Iadecola}}, \bibinfo{journal}{Phys. Rev. Lett.} \textbf{\bibinfo{volume}{123}}, \bibinfo{pages}{147201} (\bibinfo{year}{2019}), \urlprefix\url{https://link.aps.org/doi/10.1103/PhysRevLett.123.147201}.

\bibitem[{\citenamefont{Chattopadhyay et~al.}(2020)\citenamefont{Chattopadhyay, Pichler, Lukin, and Ho}}]{XY_EP}
\bibinfo{author}{\bibfnamefont{S.}~\bibnamefont{Chattopadhyay}}, \bibinfo{author}{\bibfnamefont{H.}~\bibnamefont{Pichler}}, \bibinfo{author}{\bibfnamefont{M.~D.} \bibnamefont{Lukin}}, \bibnamefont{and} \bibinfo{author}{\bibfnamefont{W.~W.} \bibnamefont{Ho}}, \bibinfo{journal}{Phys. Rev. B} \textbf{\bibinfo{volume}{101}}, \bibinfo{pages}{174308} (\bibinfo{year}{2020}), \urlprefix\url{https://link.aps.org/doi/10.1103/PhysRevB.101.174308}.

\bibitem[{\citenamefont{Yang}(1989)}]{eta_pairing_1}
\bibinfo{author}{\bibfnamefont{C.~N.} \bibnamefont{Yang}}, \bibinfo{journal}{Phys. Rev. Lett.} \textbf{\bibinfo{volume}{63}}, \bibinfo{pages}{2144} (\bibinfo{year}{1989}), \urlprefix\url{https://link.aps.org/doi/10.1103/PhysRevLett.63.2144}.

\bibitem[{\citenamefont{Yang and Zhang}(1990)}]{Hubbard_SO4}
\bibinfo{author}{\bibfnamefont{C.~N.} \bibnamefont{Yang}} \bibnamefont{and} \bibinfo{author}{\bibfnamefont{S.~C.} \bibnamefont{Zhang}}, \bibinfo{journal}{Modern Physics Letters B} \textbf{\bibinfo{volume}{04}}, \bibinfo{pages}{759} (\bibinfo{year}{1990}), \eprint{https://doi.org/10.1142/S0217984990000933}, \urlprefix\url{https://doi.org/10.1142/S0217984990000933}.

\bibitem[{\citenamefont{Vafek et~al.}(2017)\citenamefont{Vafek, Regnault, and Bernevig}}]{eta_entanglement}
\bibinfo{author}{\bibfnamefont{O.}~\bibnamefont{Vafek}}, \bibinfo{author}{\bibfnamefont{N.}~\bibnamefont{Regnault}}, \bibnamefont{and} \bibinfo{author}{\bibfnamefont{B.~A.} \bibnamefont{Bernevig}}, \bibinfo{journal}{SciPost Phys.} \textbf{\bibinfo{volume}{3}}, \bibinfo{pages}{043} (\bibinfo{year}{2017}), \urlprefix\url{https://scipost.org/10.21468/SciPostPhys.3.6.043}.

\bibitem[{\citenamefont{Mark and Motrunich}(2020)}]{extended_eta_scar}
\bibinfo{author}{\bibfnamefont{D.~K.} \bibnamefont{Mark}} \bibnamefont{and} \bibinfo{author}{\bibfnamefont{O.~I.} \bibnamefont{Motrunich}}, \bibinfo{journal}{Phys. Rev. B} \textbf{\bibinfo{volume}{102}}, \bibinfo{pages}{075132} (\bibinfo{year}{2020}), \urlprefix\url{https://link.aps.org/doi/10.1103/PhysRevB.102.075132}.

\bibitem[{\citenamefont{Shiraishi and Mori}(2017)}]{ShiraishiMori}
\bibinfo{author}{\bibfnamefont{N.}~\bibnamefont{Shiraishi}} \bibnamefont{and} \bibinfo{author}{\bibfnamefont{T.}~\bibnamefont{Mori}}, \bibinfo{journal}{Phys. Rev. Lett.} \textbf{\bibinfo{volume}{119}}, \bibinfo{pages}{030601} (\bibinfo{year}{2017}), \urlprefix\url{https://link.aps.org/doi/10.1103/PhysRevLett.119.030601}.

\bibitem[{\citenamefont{Heller}(1984)}]{scar_prl}
\bibinfo{author}{\bibfnamefont{E.~J.} \bibnamefont{Heller}}, \bibinfo{journal}{Phys. Rev. Lett.} \textbf{\bibinfo{volume}{53}}, \bibinfo{pages}{1515} (\bibinfo{year}{1984}), \urlprefix\url{https://link.aps.org/doi/10.1103/PhysRevLett.53.1515}.

\bibitem[{\citenamefont{Bogomolny}(1988)}]{scar_smoothed}
\bibinfo{author}{\bibfnamefont{E.}~\bibnamefont{Bogomolny}}, \bibinfo{journal}{Physica D: Nonlinear Phenomena} \textbf{\bibinfo{volume}{31}}, \bibinfo{pages}{169} (\bibinfo{year}{1988}), ISSN \bibinfo{issn}{0167-2789}, \urlprefix\url{https://www.sciencedirect.com/science/article/pii/0167278988900759}.

\bibitem[{\citenamefont{Kaplan and Heller}(1998)}]{scar_linearity}
\bibinfo{author}{\bibfnamefont{L.}~\bibnamefont{Kaplan}} \bibnamefont{and} \bibinfo{author}{\bibfnamefont{E.}~\bibnamefont{Heller}}, \bibinfo{journal}{Annals of Physics} \textbf{\bibinfo{volume}{264}}, \bibinfo{pages}{171} (\bibinfo{year}{1998}), ISSN \bibinfo{issn}{0003-4916}, \urlprefix\url{https://www.sciencedirect.com/science/article/pii/S0003491697957730}.

\bibitem[{\citenamefont{Kaplan}(1999)}]{scar_review}
\bibinfo{author}{\bibfnamefont{L.}~\bibnamefont{Kaplan}}, \bibinfo{journal}{Nonlinearity} \textbf{\bibinfo{volume}{12}}, \bibinfo{pages}{R1} (\bibinfo{year}{1999}), \urlprefix\url{https://dx.doi.org/10.1088/0951-7715/12/2/009}.

\bibitem[{\citenamefont{Cotler and Wei}(2023)}]{scars_qft}
\bibinfo{author}{\bibfnamefont{J.}~\bibnamefont{Cotler}} \bibnamefont{and} \bibinfo{author}{\bibfnamefont{A.~Y.} \bibnamefont{Wei}}, \bibinfo{journal}{Phys. Rev. D} \textbf{\bibinfo{volume}{107}}, \bibinfo{pages}{125005} (\bibinfo{year}{2023}), \urlprefix\url{https://link.aps.org/doi/10.1103/PhysRevD.107.125005}.

\bibitem[{\citenamefont{Radcliffe}(1971)}]{spin_coherent}
\bibinfo{author}{\bibfnamefont{J.~M.} \bibnamefont{Radcliffe}}, \bibinfo{journal}{Journal of Physics A: General Physics} \textbf{\bibinfo{volume}{4}}, \bibinfo{pages}{313} (\bibinfo{year}{1971}), \urlprefix\url{https://dx.doi.org/10.1088/0305-4470/4/3/009}.

\bibitem[{\citenamefont{Arecchi et~al.}(1972)\citenamefont{Arecchi, Courtens, Gilmore, and Thomas}}]{spin_coherent_2}
\bibinfo{author}{\bibfnamefont{F.~T.} \bibnamefont{Arecchi}}, \bibinfo{author}{\bibfnamefont{E.}~\bibnamefont{Courtens}}, \bibinfo{author}{\bibfnamefont{R.}~\bibnamefont{Gilmore}}, \bibnamefont{and} \bibinfo{author}{\bibfnamefont{H.}~\bibnamefont{Thomas}}, \bibinfo{journal}{Phys. Rev. A} \textbf{\bibinfo{volume}{6}}, \bibinfo{pages}{2211} (\bibinfo{year}{1972}), \urlprefix\url{https://link.aps.org/doi/10.1103/PhysRevA.6.2211}.

\end{thebibliography}
\end{document}